\documentclass{iau_FM}
\usepackage{graphicx}
\usepackage{natbib}
\usepackage{url}
\usepackage{amssymb}

\newcommand{\Lya}{{\rm Ly}\alpha}
\newcommand{\fesc}{f_{\rm esc}}
\newcommand{\Msun}{M_\odot}
\newcommand{\GOsaka}{\texttt{GADGET3-Osaka}}

\newcommand{\apj}{\textit{ApJ}}
\newcommand{\apjl}{\textit{ApJL}}
\newcommand{\apjs}{\textit{ApJS}}
\newcommand{\mnras}{\textit{MNRAS}}
\newcommand{\nat}{\textit{Nature}}
\newcommand{\pasj}{\textit{PASJ}}
\newcommand{\araa}{\textit{ARA\&A}}
\newcommand{\aap}{\textit{A\&A}}
\newcommand{\prd}{\textit{Phys. Rev. D}}
\newcommand{\jcap}{\textit{JCAP}}

\title[Feedback models in galaxy simulations] 
{Feedback models in galaxy simulations and probing their impact by cosmological hydrodynamic simulations}

\author[K. Nagamine]   
{Kentaro Nagamine$^{1, 2, 3}$}

\affiliation{$^1$Theoretical Astrophysics Group, Department of Earth and Space Science, \\
Graduate School of Science,  Osaka University \\ 
1-1 Machikaneyama-cho, Toyonaka, Osaka, 560-0043, Japan\\
$^2$Department of Physics \& Astronomy, University of Nevada, Las Vegas \\ 
4505 S. Maryland Pkwy, Las Vegas, NV 89154-4002, USA \\ 
$^3$Kavli IPMU, The University of Tokyo, 5-1-5 Kashiwanoha, Kashiwa, Chiba, 277-8583, Japan  \\
email: {\tt kn@astro-osaka.jp} } 

\pubyear{2023}
\jname{IAUS 373} 
\editors{T. Wong \& W.-T. Kim}
\begin{document}

\maketitle

\begin{abstract}
Feedback effects generated by supernovae (SNe) and active galactic nuclei (AGNs) are pivotal in shaping the evolution of galaxies and their present-day structures. However, our understanding of the specific mechanisms operating at galactic scales, as well as their impact on circum-galactic medium (CGM) and intergalactic medium (IGM), remains incomplete.  Galaxy formation simulations encounter challenges in resolving sub-parsec scales, necessitating the implementation of subgrid models to capture the physics occurring at smaller scales.  
In this article, we provide an overview of the ongoing efforts to develop more physically grounded feedback models. We discuss the pursuit of pushing simulation resolution to its limits in galaxy simulations and the rigorous testing of galaxy formation codes through participation in the AGORA code comparison project.  Additionally, we delve into techniques for investigating the impact of feedback using cosmological hydrodynamic simulations, specifically through $\Lya$ absorption and CGM/IGM tomography. 
Furthermore, we outline our future research directions within this field and highlight the progress made by comparing our simulation results with observational data. 
\end{abstract}

\firstsection 
\section{Introduction}

Feedback processes from supernovae (SNe) and active galactic nuclei (AGNs) have a significant impact on the regulation of galaxy formation and evolution. 
The prevailing consensus suggests that AGN feedback primarily acts to suppress star formation in massive galaxies at lower redshifts, whereas SN feedback predominantly affects star formation in low-mass galaxies at higher redshifts. 
As a result, this interplay between feedback mechanisms contributes to the observed peak in the stellar-to-halo mass relation \citep{Behroozi13}. 

\begin{figure}[t]
\begin{center}
 \includegraphics[width=5.2in]{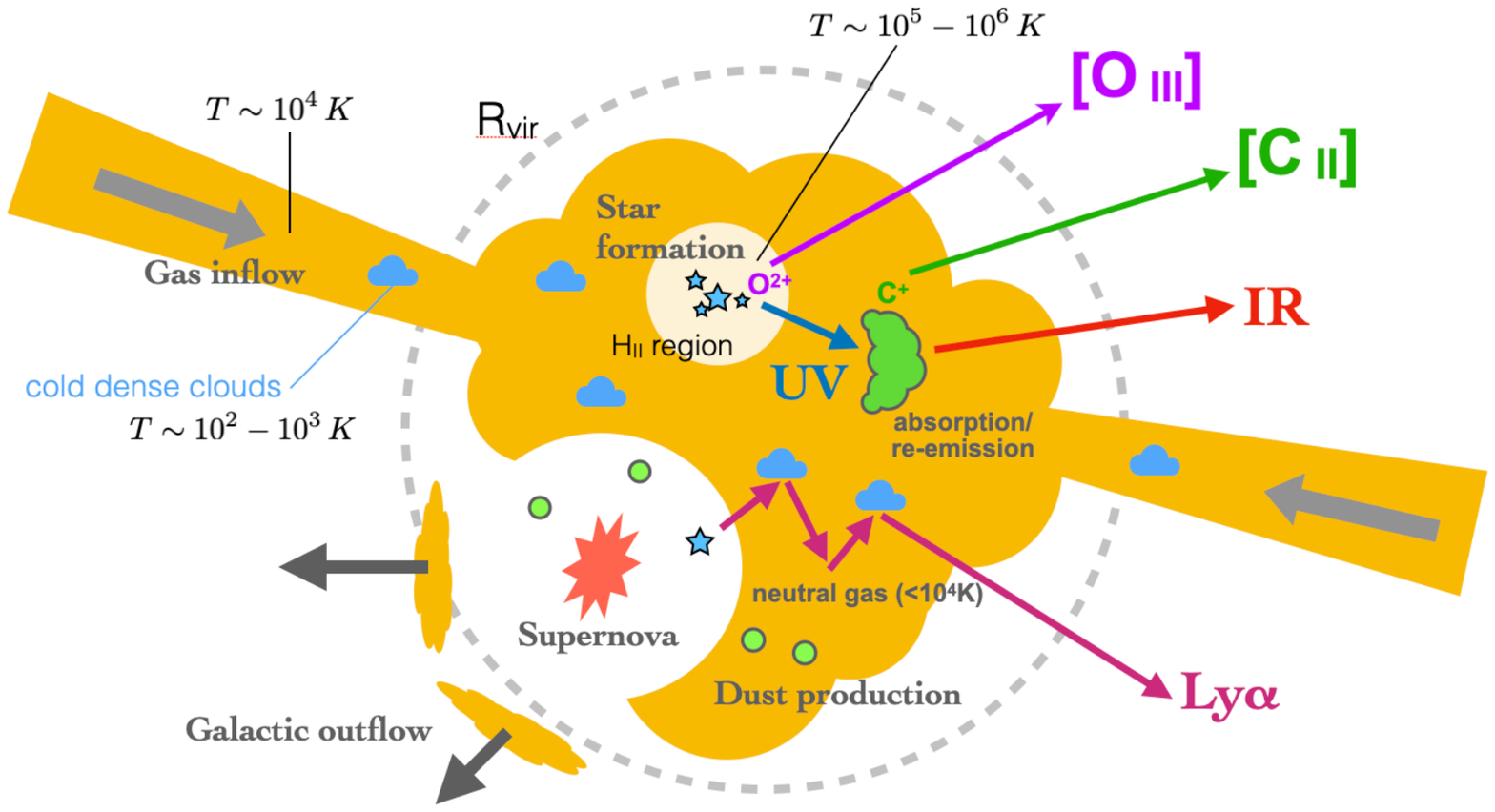} 
 \caption{Schematic illustration of key physical processes in high-redshift galaxies.  
Gas inflow occurs via cold streams with $T\sim 10^4$\,K, leading to the condensation of the gas into dense clouds  ($T\sim 10^2-10^3$\,K) through radiative cooling.  
Subsequent star formation gives rise to massive stars that ionize the surrounding ISM, creating {\sc H ii} regions. 
The UV radiation from massive stars impinges on the ISM surface, leading to the formation of photo-dissociation regions (PDRs),  which emit infrared lines like [{\sc C ii}] as reprocessed radiation. 
These emissions from high-$z$ galaxies have been computed using cosmological zoom-in hydrodynamic simulations \citep[e.g.,][]{Arata19,Katz22a,Pallottini22}.  
In addition, prominent $\Lya$ emissions are also observed from high-$z$ galaxies.
After several million years, the massive stars die, resulting in the ejection of gas as galactic outflows.
 }
   \label{fig1}
\end{center}
\end{figure}

High-redshift galaxies serve as excellent testbeds for examining feedback models due to the numerous intriguing physical processes taking place within them. 
In Figure~\ref{fig1}, we present a schematic diagram encapsulates some of these processes. 
At high redshifts, low-metallicity gas streams into dark matter halos through narrow, cold flows, providing ample fuel for star formation \citep{Keres05,Wright21}.  
However, as we transition towards lower redshifts ($z\lesssim 2$), gas accretion shifts to the hot mode, and prominent cold streams diminishes in simulations \citep{Faucher11,Nelson16}.  
The cosmic star formation rate density (SFRD) exhibits a broad peak around $z\simeq 3-5$ \citep[e.g.,][and references therein]{Nag00,Nag04d,Nag06c,Kistler09,Madau14}, with the rising SFRD at high redshifts driven by the active formation of dark matter halos and galaxies through gravitational instability \citep{Schaye10}.  
Observations employing instruments like ALMA have provided valuable insights into the emission lines expected from high-redshift galaxies, including Ly$\alpha$, [{\sc C ii}], [{\sc O iii}] lines  \citep[e.g.,][]{Smit18,Hashimoto18,Hashimoto19}. 
These observations suggest the presence of very early onset of star formation at $z\sim 15$. 

The escape fraction ($\fesc$) of ionizing and ultraviolet (UV) photons stands as a critical physical parameter that profoundly influences the radiative characteristics of high-redshift galaxies. However, estimating $\fesc$ observationally is challenging, and only a limited number of approximate measurements have been made. Hence, it is desirable to directly predict $\fesc$ for different types of galaxies using hydrodynamic simulations of galaxy formation  \citep[e.g.,][]{Cen03d,Razoumov06a,Gnedin08b,Wise09,Yajima17}. 
To make reliable predictions of $\fesc$, accurate computations of interstellar medium (ISM) structure and, thus, the attainment of high resolution are indispensable. 
Furthermore, understanding the intricate details of $\fesc$ holds significant importance in determining whether the reionization of the universe favors the "Early" or "Late" scenarios \citep[e.g.,][]{Finkelstein19,Naidu20}. 
By conducting in-depth studies on the escape fraction, we can gain valuable insights into the processes contributing to the ionization state of the universe and the timing of reionization events. 
Such investigations provide crucial information for comprehending the complex interplay between galaxies, their radiation, and the overall evolution of the early universe.

\section{Feedback models in galaxy formation simulations}

In early cosmological hydrodynamic simulations, SN feedback was often simplified by injecting thermal energy on large scales ($>$\,kpc) \citep{Cen92a,Katz92a,Katz96a,Cen99a}.  However, this approach faced challenges in low-resolution simulations, as the injected thermal energy would rapidly dissipate through radiation due to the inability to resolve the detailed Sedov-Taylor phase of each SN or collective superbubble.  
This issue, known as the overcooling problem, posed a significant hurdle. 
To overcome the overcooling problem, effective models of SN feedback have been developed, employing various strategies within galaxy formation simulations.  These strategies include: 
(i) Ignoring and bypassing unresolved scales; (ii) Scaling up the energy dynamics to a resolvable scale by considering cumulative energies; or (iii) Modeling physics on unresolved scales via subgrid models. 

As examples of method (i), the ``delayed cooling" model temporarily ignores the cooling after a supernova event to enhance the impact of  thermal feedback \citep{Thacker00a,Stinson06}.  The ``constant velocity wind" model of \citet{Springel03b} stochastically kicks gas particles and disables hydro forces until the wind particles exit the galaxies, within a smoothed particle hydrodynamics (SPH) code. 
In method (ii), the ``stochastic thermal feedback" model increases the temperature of neighboring fluid elements by a certain value, $\Delta T$ \citep{Kay03,Vecchia12}, so that the subsequent evolution of the hot bubble can be solved by a hydro solver with efficient thermal feedback.  However, the choice of $\Delta T$ remains somewhat arbitrary and uncertain. 
Method (iii) includes the ``multiphase ISM model," where a single SPH particle is treated as a multiphase gas, and energy exchange between the hot and cold phases is accounted for using a subgrid equilibrium model \citep{Yepes97,Springel03b,Keller14}.
Another approach, as part of method (iii), involves injecting the terminal momentum of a single SN explosion based on the Sedov--Taylor solution \citep{Kimm14,Hopk18}. 
For further discussions on different feedback treatments, refer to \citet{Nag18} and \citet{Oku22}.

\begin{figure}[t]
\begin{center}
 \includegraphics[width=5.2in]{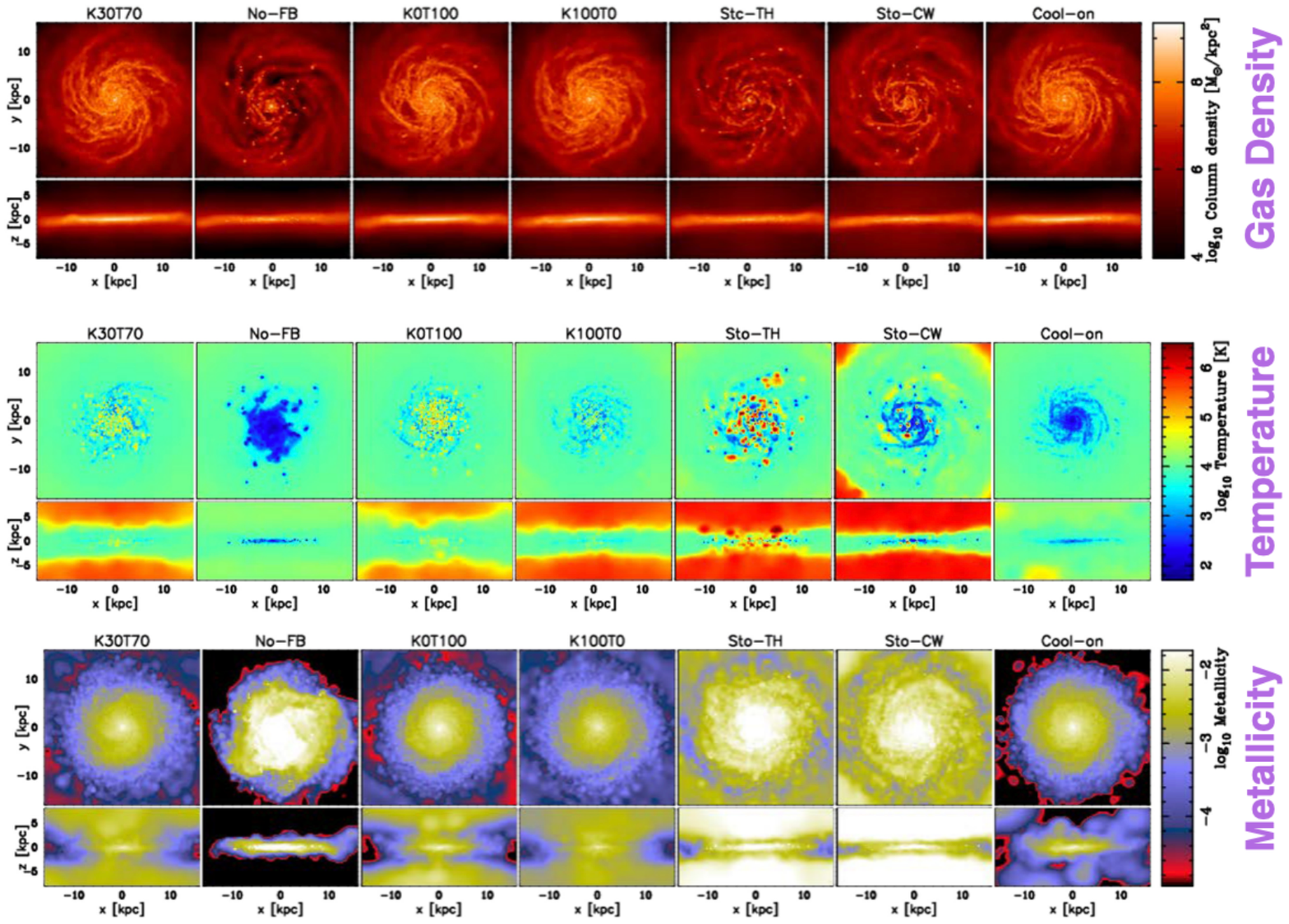} 
 \caption{An example of testing SN feedback models in isolated galaxy simulations, showing the projected gas density, the mass-weighted temperature, and the metallicity from top to bottom rows, respectively \citep{Shimizu19}.  Within each row, the top panels show a face-on view, while the lower panel presents an edge-on view. 
  }
   \label{fig2}
\end{center}
\end{figure}

In \citet{Shimizu19}, we employed a combination of the delayed cooling model and kinetic feedback using the Sedov-Taylor self-similar solution within the \GOsaka\ code. 
Figure~\ref{fig2} presents an example of testing different SN feedback models. 
In the fiducial model (K30T70), we injected 30\% of the SN energy as kinetic energy and 70\% as thermal energy, following approaches in previous studies \citep[e.g.,][]{Chevalier74,Durier12}.  
To ensure the effectiveness of thermal feedback, cooling was temporarily disabled for the neighboring particles that received the thermal feedback energy.  
The gas distribution image in the face-on view of the "No-feedback" (No-FB) run exhibits a highly clumpy, cold, and dense gas distribution, with a noticeable absence of hot gas above the disk.  As the fraction of thermal energy injection increases, the spiral arms become more diffuse compared to the Fiducial run.
The ``Stochastic thermal" (Stc-TH) and ``Stochastic constant wind velocity" (Sto-CW) runs also display clumpy knots within the disk.  However, the feedback effects above the disk are more pronounced in these runs, resulting in the presence of hot outflowing gas that excessively enriches the CGM. 
When cooling is activated in the "Cool-on" run, the gas heated by feedback rapidly cools down and remains confined within the disk. Consequently, the Cool-on run exhibits similar disk features to the No-FB run.
The Osaka feedback model presented in \citet{Shimizu19} offers a valuable comparison of various SN feedback treatments. 
This model successfully achieves self-regulation of star formation and naturally generates galactic outflows.
However, it still contained some unphysical treatments, such as the temporary disabling of cooling for effective thermal feedback. 

In \citet{Oku22}, we revisited the concept of single SN remnant  (SNR) and superbubble, drawing inspiration from earlier studies \citep{Chevalier74,Weaver77,Tomisaka86,Ostriker88,Martizzi15,KimCG15,KimCG17b}, and investigated the metallicity dependence of the terminal moment of the SN shell. 
Using the Eulerian hydrodynamic code \texttt{ATHENA++}, we extended the analytic solution for the SNR shell-formation time by \citep{KimCG15} to incorporate the effect of metallicity.  Additionally, we obtained an analytic solution for the formation time of the superbubble shell.   
We found a universal scaling relations for the temporal evolution of momentum and radius for a superbubble when scaled by their values at the shell-formation time.
Building upon these findings, we developed a SN feedback model based on the \texttt{ATHENA++} simulation results.  This involved employing Voronoi tessellation around each star particle and integrating it into the \GOsaka\ SPH code. 
We examined the mass/energy/metal loading factors and found that our stochastic thermal feedback model generated galactic outflows capable of transporting metals above the galactic plane while exhibiting a modest suppression of star formation.  
Incorporating additional mechanical feedback further suppressed star formation and improved the agreement between simulation results and observations of the Kennicutt--Schmidt relation, considering the uncertainties in the observed data.  
We argued that both thermal and mechanical feedback are crucial in the SN feedback model of galaxy evolution, particularly in SPH simulations where individual SN bubbles remain unresolved. 

Some simulations have made significant progress in pushing the resolution limits by focusing on low-mass galaxies. 
For example, \citet{Hu19} employed the GADGET-3 SPH simulation to study SN feedback in a dwarf galaxy residing in a dark matter halo with a virial mass of $M_{\rm vir}=10^{10}\,\Msun$.  They achieved a remarkable resolution of $m_{\rm gas} = 1\,\Msun$, along with 0.3\,pc for gravitational softening length and SPH smoothing length.  
Despite this high resolution, certain assumptions were still necessary, such as determining the fraction of energy given as kinetic energy and determining the number of SPH particles that receive the feedback energy. 
Nonetheless, they successfully simulated the formation of superbubbles with sizes of a few hundred parsecs and investigated their breakout from the galactic disk. 
Similarly, \citet{Ma20} conducted simulations of dwarf galaxy formation during the reionization epoch using cosmological zoom-in simulations with the {\small FIRE-2 GIZMO}. 
They focused on a halo with a virial mass of $M_{\rm vir}=3.7\times 10^{10}\,\Msun$ and achieved a mass resolution of $m_{\rm gas} = 100\,\Msun$ and a spatial resolution of approximately one parsec. 
They examined the location of star formation within these high-redshift dwarf galaxies in relation to the superbubbles walls and observed significant spatio-temporal variations in the escape fraction of ionizing photons. 
These findings are in line with previous studies by \citet[cf.][]{Wise09,Kimm14}. 
 As simulations continue to improve in mass resolution, approaching a "star-by-star" level, it becomes necessary to devise methods for stochastically sampling the initial mass function (IMF) for the star formation model. Several studies, including \citet{Ploeckinger14,Hu19,Hirai21}, have explored this approach, but further investigation is needed to understand its impact on feedback processes as well as the overall galaxy evolution.

\section{The AGORA code comparison project}

Conducting code comparison projects is indeed an important approach to test and improve galaxy formation codes. 
Several notable projects have been undertaken in this regard, including the Santa Barbara cluster comparison project
 \citep{Frenk99}, the Aquila project \citep{Scannapieco12}, the nIFTy project \citep{Knebe15}, and the AGORA project \citep{Kim14,Kim16,Fabrega21}.  
 These initiatives bring together different research groups to systematically compare their simulation results, exchange ideas, identify strengths and weaknesses, and foster improvements in galaxy formation modeling.
 
The Santa Barbara cluster comparison project played a crucial role in highlighting the diverse results obtained from different hydrodynamic schemes and the issue of spurious entropy generation in SPH codes.  
This project had a profound impact on the development of new SPH schemes that could better resolve shocks, such as the density-independent scheme \citep{Saitoh13} and more general formulations based on Lagrangian-based derivations  \citep{Springel02,Hopk13b}.  
These improved SPH schemes offer enhanced stability and shock resolution compared to traditional versions.
The incorporation of these new schemes has led to significant improvements in the fidelity of SPH simulations, addressing some of the earlier challenges and limitations associated with spurious entropy generation.

The Aquila comparison project focused on investigating code-to-code variations in galactic properties at $z=0$, including stellar mass, size, morphology, and gas content. The project concluded that, due to different feedback prescriptions employed in the simulations, the models were not yet capable of uniquely predicting galactic properties, even when the assembly history of dark matter halos was the same. This highlights the importance of refining and calibrating feedback models to achieve more accurate and consistent predictions.

\begin{figure}[t]
\begin{center}
 \includegraphics[width=5.2in]{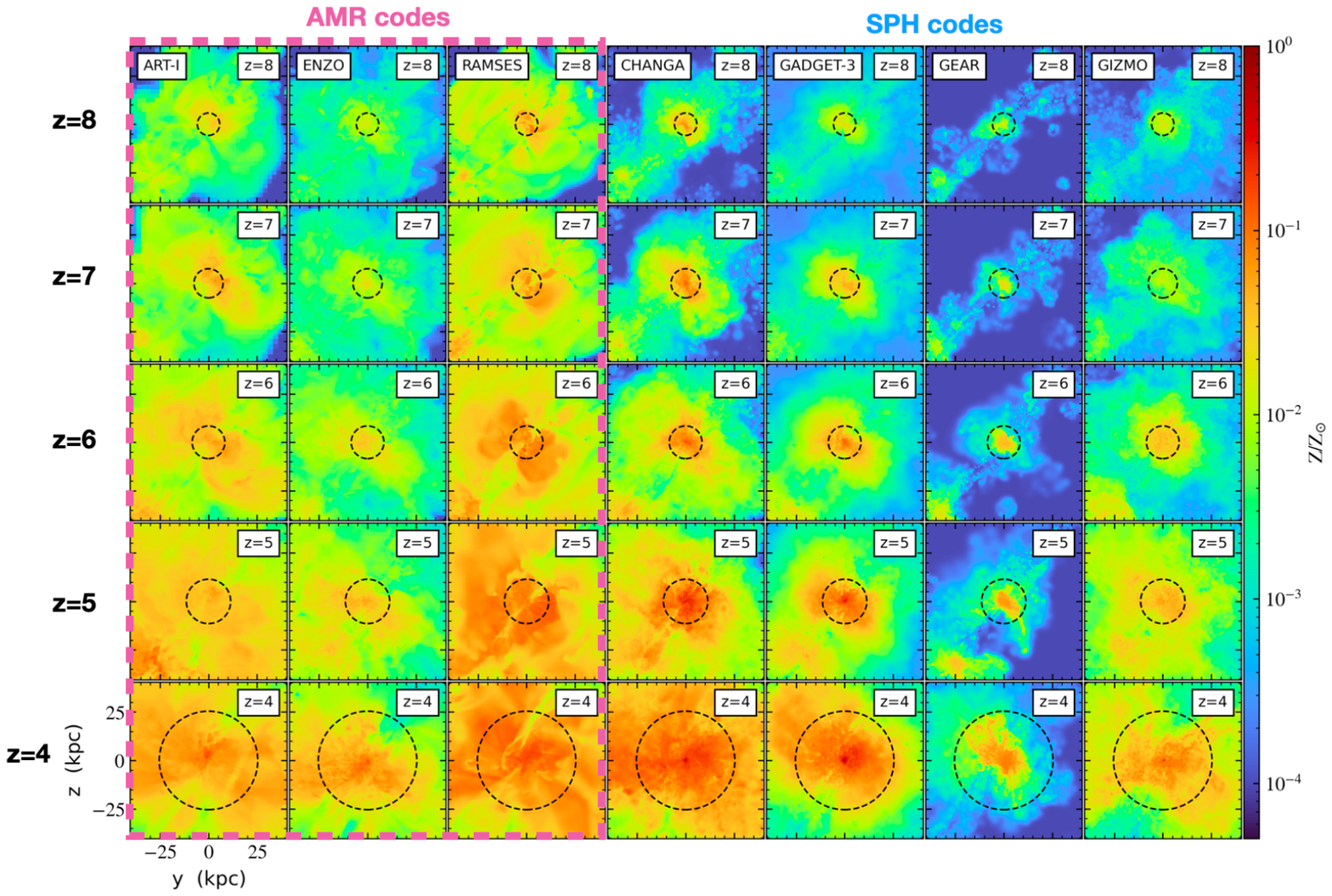} 
 \caption{Comparison of projected (density-square-weighted) metallicity of seven different code from $z=8$ to $z=4$.  Adapted from Fig.\,18 of \citep{Fabrega21}.   }
   \label{fig3}
\end{center}
\end{figure}

The AGORA project (Assembling Galaxies of Resolved Anatomy)  (\footnote{\url{https://sites.google.com/site/santacruzcomparisonproject/}}) was designed to enable a more controlled environment for galaxy formation simulations. The project brought together multiple simulation codes, encompassing SPH, AMR, and moving mesh methods, and aimed to achieve consistent results by implementing common astrophysics setups.
To establish a common baseline for comparison, the AGORA project initiated rigorous calibration steps that included adopting a common star formation recipe and utilizing the same Grackle cooling module \citep{Smith17} across all participating codes.  

With these standardized astrophysics setup, the AGORA project demonstrated more consistent behaviors among the different  simulation codes.  \citet{Kim16} concluded that modern high-resolution galaxy formation simulations are primarily influenced by the input physics, such as feedback prescriptions, rather than intrinsic differences in numerical schemes. 

Building upon these findings, \citet{Fabrega21} extended the comparison to cosmological zoom-in hydro simulations, 
and seven contemporary astrophysical simulation codes ({\small ART-I, ENZO, RAMSES, CHANGA, GADGET-3, GEAR, and GIZMO}) were compared.  
The comparison process involved four systematic calibration steps, starting from a simple adiabatic run without cooling and star formation, and gradually incorporating cooling, heating, and star formation in subsequent steps.  
In the final step, each code was tasked to reproduce a stellar mass of $\sim 10^9\,\Msun$ at $z=4$ within a halo that would grow to  $10^{12}\,\Msun$ by $z=0$, employing code-specific SN feedback recipes. 

With a physical resolution of $\lesssim$\,100\,pc at $z=4$, the participating codes demonstrated a general agreement on gas and stellar properties.  However, interesting differences emerged in the temperature and chemical enrichment of the CGM due to variations in feedback treatments, as illustrated in Fig.~\ref{fig3}, and Strawn et al. (2023, in preparation). 
These results emphasized the need for further refinement and constraint of SN and AGN feedback models through comprehensive comparisons with a wide range of observations.

Overall, the AGORA project provides a valuable framework for comparing simulation codes, promoting a deeper understanding of their similarities, differences, and areas of improvement. It highlighted the importance of standardized astrophysics setups and the ongoing development of more accurate and constrained feedback models for advancing our understanding of galaxy formation and evolution.

\section{Probing the impact of feedback by cosmological simulations}

It is crucial to complement studies of feedback in isolated galaxies and zoom-in simulations with large-scale cosmological hydrodynamic simulations. These simulations offer the advantage of larger box sizes and a broader sample of galaxies, enabling researchers to investigate important galaxy statistics such as the galaxy stellar mass/luminosity functions and the stellar-to-halo-mass ratio (see the contribution by R. Somerville in these proceedings).

In addition to galaxies, we would also like to probe the distribution of diffuse baryons via absorption and emission lines. 
For example, the distribution of neutral hydrogen ({\sc Hi}) probed by the $\Lya$ forest \citep[e.g.,][]{Weymann81,Cowie95,Rauch98} reflects the strength of UV background radiation field and the local ionizing radiation. 
We anticipate that the impact of feedback is imprinted in CGM/IGM \citep{Cen94,Hernquist96,Miralda96,Zhang97,Zhang98,Theuns02,Cen05,Kollmeier06}. 
The $\Lya$ forest serves as a powerful tool for cosmological studies and has been used to constrain cosmological parameters and the matter power spectrum
\citep{Weinberg98a,Croft98,McDonald06,Irsic17a}, as well as to investigate the mass of warm dark matter particles or neutrinos \citep[e.g.][]{Viel05,Viel13a,Palanque15b}.
This line of research is also related to the unresolved `Missing baryon problem' \citep{Cen99a,Nicastro05,Shull12,deGraaff19}, 
which refers to the challenge of observationally accounting for the entire cosmic baryon content.

\begin{figure}[t]
\begin{center}
  \includegraphics[width=2.4in]{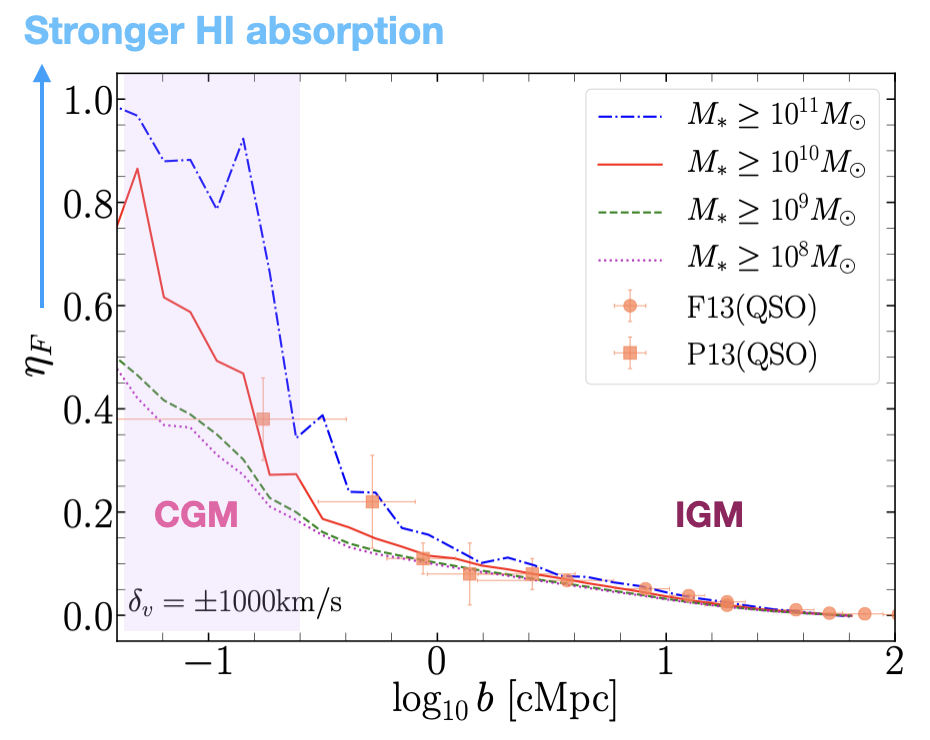} 
 \includegraphics[width=2.8in]{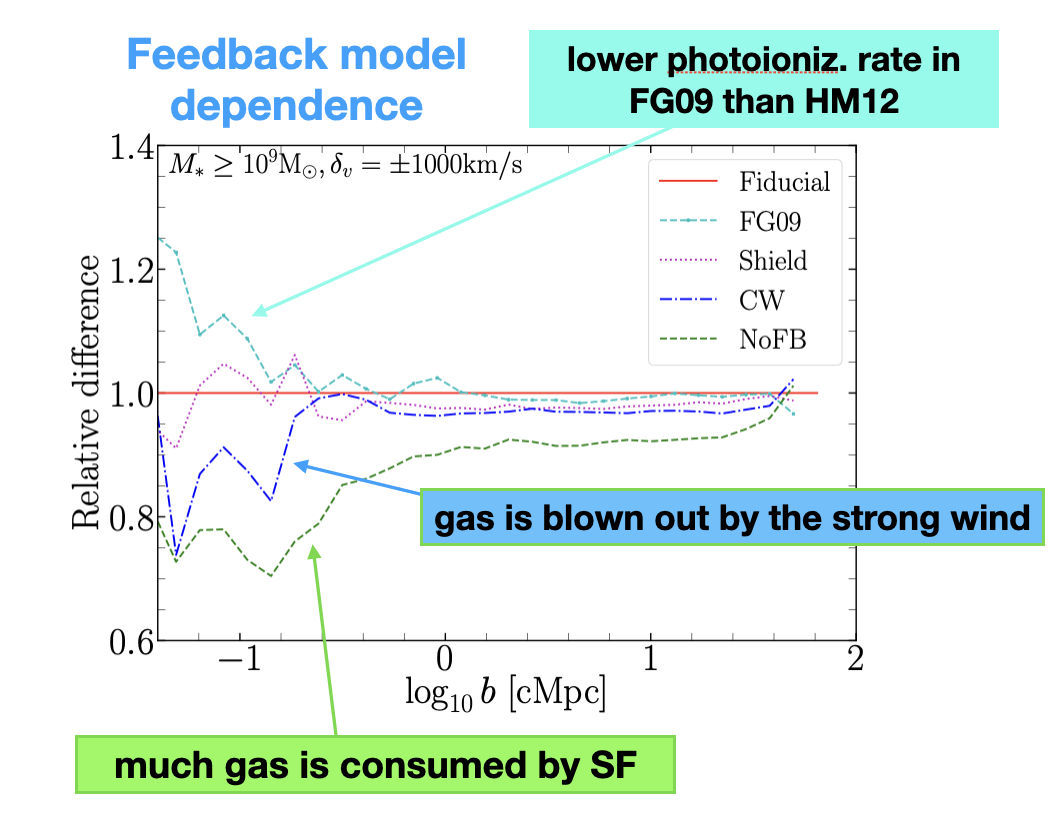} 
 \caption{{\it Left:} Ly$\alpha$ flux contrast as a function of impact parameter from nearby galaxies.  The data points are from \citet[][orange filled circle; F13]{Font-Ribera13} and \citet[][orange filled square; P13]{Pro13}.
 {\it Right:} Relative difference in the flux contrast from the Fiducial model, showcasing runs with different feedback and UVB treatments.  Both figures are adapted from \citet{Nag21}.   }
   \label{fig4}
\end{center}
\end{figure}

IGM tomography represents an enhanced version of the $\Lya$ forest technique, enabling the generation of three-dimensional contour maps of {\sc Hi} density in the IGM. This advanced approach utilizes a larger sample of star-forming galaxies as background sources, in addition to quasar sight-lines, to reconstruct the spatial distribution of {\sc Hi} gas in three dimensions. Several groups have already demonstrated the feasibility of this approach \citep[e.g.,][]{Lee14b,Lee18,Cai16,Mukae20},  
and  massive protoclusters at redshifts $z=2-3$ have been identified \citep{Lee16a,Cai17}.  
Moreover, this technique enables the derivation of the correlation between galaxy overdensity and {\sc Hi} overdensity, providing valuable insights into the relationship between galaxies/AGNs and the surrounding {\sc Hi} gas \citep{Mukae17,Liang21,Momose21a}. 

The scientific objectives of IGM tomography are: (i) to characterize the cosmic web at $z>2$, (ii) to study the association between galaxies/AGNs and {\sc Hi} gas, and (iii) to identify protoclusters and voids in an {\it unbiased} manner.

As a pathfinder to the IGM tomography studies by the Subaru PFS \citep{Takada14,Greene22} and upcoming observations by the JWST/TMT/ELT,
\citet{Nag21} investigated the impact of feedback on basic $\Lya$ forest statistics 
by creating a light-cone data set at $z=2-3$ and generating a mock $\Lya$ forest data.
They used five cosmological hydro simulations by \GOsaka\ code with different models of feedback and UVB treatment (comoving boxsize $L_{\rm box}$=147.6\,Mpc, particle number $N=2\times 512^3$),  and examined the 1D flux probability distribution function, 1D flux power spectrum, flux contrast vs. impact parameter from galaxies, and {\sc Hi}--galaxy cross-correlation.
The flux contrast is defined as $\eta_F \equiv - \delta_F = 1-\frac{F}{\langle F \rangle}$, where $F$ is the transmitted flux ($F = e^{-\tau}$), and  ${\langle F \rangle}$ is the average effective $\Lya$ optical depth adjusted to the observed value \citep{Becker13b}. 
Higher $\eta_F$ in the vicinity of galaxies means stronger absorption, i.e., more {\sc Hi} (left panel of Fig.\,\ref{fig4}).
In other words, they found stronger {\sc Hi} absorption with decreasing impact parameter from galaxies, consistently with earlier simulation results \citep[e.g.,][]{Bruscoli03,Kollmeier03,Kollmeier06,Meiksin15,Turner17,Meiksin17,Sorini18}. 
Their simulation results demonstrated overall agreement with current observational data, but with some interesting discrepancies  of about  30\% on small scales that are due to different treatments of feedback and UVB, or varying observational conditions (right panel of Fig.\,\ref{fig4}).   
The massive galaxies with $M_{\star} \ge 10^{10}\,\Msun$  strongly contribute to the flux contrast signal (left panel of Fig.\,\ref{fig4}), while lower-mass galaxies in the range of $M_\star \approx 10^8 - 10^{10}\,\Msun$ dilute the flux contrast signal from massive galaxies when averaged over the entire galaxy sample.  
The variations in $\eta_F$ on scales of $<1$\,Mpc can be probed with future IGM tomography surveys with dense background source sampling by JWST/ELT/TMT. 
On larger scales, the average flux contrast smoothly connects to the IGM level, supporting the spherical infall model and concordance $\Lambda$ cold dark matter model, as also found by \citet{Meiksin17,Sorini18}. 
Interestingly, \citet{Sorini20} reported negligible impact of AGN feedback on the flux contrast, suggesting that stellar feedback primarily determines the average physical properties of CGM at $z=2-3$.  However, further investigation in simulations incorporating AGN feedback is warranted to confirm this finding \citep[cf.][]{Tillman22}.

In addition to {\sc Hi} distribution, metal distribution can also be probed by emission and absorption lines.  For example, the MEGAFLOW project has observed Mg\,{\sc ii} lines in both absorption and emission in the galactic wind region of $z\sim 0.7$ galaxy \citep{Zabl20,Zabl21}. 
In a different study, \citet{Nelson21} computed the resonantly scattered Mg\,{\sc ii} emission from the TNG50 simulation. 
Their analysis indicates that the simulated galaxies exhibit somewhat steeper profiles (i.e., a faster decline with increasing radii) compared to the observed data points (see their Fig.\,3).  However, the currently observed sources are especially bright ones that are easily detectable, therefore, further comparisons with lower mass systems are necessary in the future. 

Interestingly, similar trends are observed in other emission lines, such as [{\sc C ii}]  \citep{Arata20,Fujimoto19}, and $\Lya$ \citep{Zhang20}, where the simulated galaxies fail to reproduce the observed extended emission profiles.  
These discrepancies have interesting implications for the feedback efficiencies, such as the mass-loading factor of metals in galactic outflows \citep{Pizzati20}, and therefore warrant further studies to constrain the efficiencies of chemical enrichment in CGM and IGM.

\section{Summary}

In this article, we reviewed various feedback treatments in galaxy simulations, and discussed our development of physically-based SN feedback models and the tests within the AGORA project using isolated galaxies and zoom-in cosmological hydrodynamic simulations. 
We argued that considering both thermal and kinetic modes of SN feedback is important at the current resolution level ($\gtrsim$\,10\,pc).  
In our recent work, \citet{Oku22} showed that the kinetic feedback suppresses star formation while stochastic thermal feedback drives strong metal outflows.
Further studies on both small and large scales are crucial to fully understand the role of feedback in galaxy evolution and the chemical enrichment of CGM/IGM. 

Notably there are indications that very high-resolution simulations ($\lesssim$\,pc scale) exhibit weaker winds compared to  larger-scale simulations that capture the physics of galactic winds on supergalactic scales (see contributions by E. Ostriker and C.-G. Kim in this proceedings, as well as \citealt{Hu19}).  
Additionally, the inclusion of additional physics, such as cosmic rays and magnetic fields, will be essential for constructing more physically plausible models of star formation and feedback \citep[e.g.,][]{Hopk22}.

It might also be possible to constrain the physics of feedback at larger scales of circumgalactic and intergalactic scales utilizing the $\Lya$ absorption by neutral hydrogen (commonly referred to as IGM tomography) and distribution of metals and dust. 
For example, \citet{Nag21} have shown that SN feedback influences the radial distribution of {\sc H i} gas and the $\Lya$ flux contrast signal at $\sim$30\% level. 
Future comparisons between simulations and the CGM/IGM tomography surveys by WEAVE, MOONS, Subaru PFS, JWST, ELT, and TMT will provide valuable insights and further constrain the physics of feedback.

\vspace{5mm}
I am grateful to all of my recent collaborators on the research results discussed in this article, including S. Arata, R. Cen, K. G. Lee, R. Momose, Y. Oku, L. Romano, I. Shimizu, K. Tomida, H. Yajima, and everyone in the AGORA project. 


\end{document}